\documentclass[aps,pra,twocolumn,showpacs,groupedaddress,nofootinbib]{revtex4-1}
\usepackage{bm,amsmath}
\usepackage{graphicx}
\usepackage{afterpage}
\usepackage{amssymb}
\usepackage{amsmath}
\usepackage{txfonts}
\usepackage{epsfig}
\usepackage{natbib}
\usepackage{color}
\usepackage{bm}
\usepackage[pdftex,colorlinks,linkcolor=blue,citecolor=blue,urlcolor=blue]{hyperref}
\usepackage{array} 
\usepackage{booktabs}
\usepackage[normalem]{ulem}
\usepackage{color}
\newcommand{\comment}[1]{}

\definecolor{gray}{gray}{0.6}

\begin{document}
\title{Beyond Thomas--Fermi analysis of the density profiles of a miscible two-component Bose--Einstein condensate}
\author{J. Polo$^{\ast}$}
\affiliation{Departament de F\'{\i}sica, Universitat Aut\`{o}noma de Barcelona, E-08193 Bellaterra, Spain}
\email[]{Juan.Polo@uab.cat}
\author{P. Mason}
 \affiliation{School of Physics and Astronomy, University of Birmingham, Edgbaston, Birmingham B15 2TT, United Kingdom}
\author{S. Sridhar}
\affiliation{Joint Quantum Centre (JQC) Durham--Newcastle, Department of Physics, Durham University, Durham DH1 3LE, United Kingdom} 
\author{T. P. Billam}
\affiliation{Joint Quantum Centre (JQC) Durham--Newcastle, Department of Physics, Durham University, Durham DH1 3LE, United Kingdom} 
\author{V. Ahufinger}
\affiliation{Departament de F\'{\i}sica, Universitat Aut\`{o}noma de Barcelona, E-08193 Bellaterra, Spain} 
\author{S. A. Gardiner}
\affiliation{Joint Quantum Centre (JQC) Durham--Newcastle, Department of Physics, Durham University, Durham DH1 3LE, United Kingdom} 
\date{\today}
\begin{abstract}
We investigate a harmonically trapped two-component Bose--Einstein condensate within the miscible regime, close to its boundaries, for different ratios of effective intra- and inter-species interactions. We derive analytically a universal equation for the density around the different boundaries in one, two and three dimensions, for both the coexisting and spatially separated regimes. We also present a general procedure to solve the Thomas--Fermi approximation in all three spatial dimensionalities, reducing the complexity of the Thomas--Fermi problem for the spatially separated case in one and three dimensions to a single numerical inversion. Finally, we analytically determine the frontier between the two different regimes of the system.
\end{abstract}
\maketitle
%%%%%%%%%%%%%%%%%%%%%%%%%%%%%%%%%%%%%%%%%%%%%%%%%%%%%%%%%%%%%%%%%%%%%%%%%%%%%%%%%%%%%%%%%%%%%%%%%%%%%%%%%%%%%%%%%%%%
%%%%%%%%%%%%%%%%%%%%%%%%%%%%%%%%%%%%%%%%%%%%%%%%%%%%%%%%%%%%%%%%%%%%%%%%%%%%%%%%%%%%%%%%%%%%%%%%%%%%%%%%%%%%%%%%%%%%
\section{Introduction}
\label{sec:Introduction}
The experimental realisation of Bose--Einstein condensation (BEC) in dilute atomic vapours was a landmark achievement of late twentieth century physics \cite{Cornell_1995,Ketterle_1995,Hulet_1995}, and there are now many elements that can be cooled down to quantum degeneracy \cite{Hulet_1995,Cornell_1995,Ketterle_1995,Hulet_1997,Thomas_1998,Wieman_2000,Aspect_2001,Inguscio_2001,Lev_2011,Ferlaino_2012}. A significant number of two-component Bose--Einstein condensates (TCBEC) have also been reported, as mixtures of two atomic species \cite{Inguscio_2002,Simoni_2002,Inguscio_2008,Cornish_2011}, two isotopes of the same species \cite{Wieman_2008}, or as two hyperfine states of the same isotope \cite{Wieman_1997,Ketterle_1998,Cornell_1998_1,Cornell_1999,Inguscio_2000,Aspect_2001_1,Cornell_2004,Hall_2009}.

Many theoretical studies have addressed the density profiles of TCBECs depending on the ratio between the intra- and inter-species interaction strengths \cite{Shenoy_1996,Bohn_1997,Bigelow_1998,Chui_1998,Modugno_2002}. However, most of these theoretical studies are numerical or based on the Thomas--Fermi (TF) approximation. This approximation, introduced for the single component case \cite{Thomas_1927,Fermi_1927}, describes the basic features of the ground state of a BEC with large interatomic interactions. The TF approach neglects the kinetic energy term in the time independent Gross--Pitaevskii equation, on the grounds that its contributions are to a significant extent dominated over by those due to the nonlinear interaction term.  It can give good approximations, for instance,  of the condensate chemical potential or of the order parameter near its maximum value. However, close to the order parameter boundaries, where the atomic densities are low, the TF approximation cannot provide the condensate density profile. Knowing the wave function of the condensate around these boundaries is very important to characterize for instance the actual kinetic energy \cite{Stringari_1996,Feder_1998}, the tunneling rate across a potential barrier \cite{Stringari_1996}, or in the case of TCBEC systems, the penetration of one component into the other \cite{Chui_1998}. Several works have proposed new analytical approximations beyond the TF approach for single component BECs \cite{Stringari_1996,Feder_1998,Theodorakis_2004,Delgado_2007,Salman_2012} and for the two component case in the immiscible regime \cite{Chui_1998}. Here we present a new analytical approach to study the density profile of TCBECs within the miscible regime, around the regions where the TF approach fails, by deriving a universal equation. We also introduce a general procedure to solve the TF approximation of TCBECs in one (1D), two (2D) and three dimensions (3D) and we provide an analytical formula that determines the frontier between the different regimes of the system \cite{Shenoy_1996}. Our method also reduces the complexity of the numerical inversion required in the TF approach for the one and three dimensional cases \cite{Tsubota_2001,Modugno_2002} for the spatially separated regime.

The paper is organized as follows. In Sec.~\ref{sec:TF} we present the equations that describe TCBECs and the general form of the two-component TF approximation in the particular case of an isotropic harmonic potential. In Sec.~\ref{sec:BTF} we derive a universal equation governing the behavior of the density profile close to the different boundaries of the system. Finally, in Sec.~\ref{sec:TFG} we develop a procedure to solve the two-component TF approach in a general way. We compare our approach, for each dimensionality, with the numerical solution of the coupled Gross--Pitaevskii equation in Sec.~\ref{sec:results}, and present our conclusions in Sec.~\ref{sec:conclusions}.
%%%%%%%%%%%%%%%%%%%%%%%%%%%%%%%%%%%%%%%%%%%%%%%%%%%%%%%%%%%%%%%%%%%%%%%%%%%%%%%%%%%%%%%%%%%%%%%%%%%%%%%%%%%%%%%%%%%%%%%%%%
\section{Ground-state of a Two-component Bose--Einstein condensate}
\label{sec:TF}
%%%%%%%%%%%%%%%%%%%%%%%%%%%%%%%%%%%%%%%%%%%%%%%%%%%%%%%%%%%%%%%%%%%%%%%%%%%%%%%%%%%%%%%%%%%%%%%%%%%%%%%%%%%%%%%%%%%%%%%%%%
\subsection{Gross--Pitaevskii equations and the Thomas--Fermi limit}
The ground state of a TCBEC at zero temperature within the mean-field approximation is typically well described by the time-independent two-component Gross--Pitaevskii equations (TCGPEs):
\begin{equation}
\label{GPE1}	\left(-\frac{\hbar^{2}\nabla^{2}}{2m_{s}}+V_{s}+\tilde{g}_{s}N_{s}|\Psi_{s}|^{2}+\tilde{g}_{12}N_{3-s}|\Psi_{3-s}|^{2}-\mu_{s}\right)\Psi_{s}=0\;,
\end{equation}
where $s=1\text{ or }2$ refers to each component of the BEC, whilst $m_{s}$, $N_{s}$, $V_{s}(\mathbf{r})$ and $\mu_{s}$ are the mass, number of atoms, external potential and chemical potential of the $s$ component, respectively. The intra- and inter-species interaction coefficients are given by $\tilde{g}_{s}>0$ and $\tilde{g}_{12}$, respectively. We will assume that the considered TCBEC is formed by atoms of the same species in two different spin states \cite{Wieman_1997,Ketterle_1998,Cornell_1998_1,Cornell_1999,Inguscio_2000,Aspect_2001_1,Cornell_2004,Hall_2009}. This means that we can set $m_{1}=m_{2}=m$.  For simplicity, we consider equal trapping potentials $V_{1}(\mathbf{r})=V_{2}(\mathbf{r})=V(\mathbf{r})$ which in our case will be isotropic and harmonic. Nevertheless, our results can be straightforwardly generalized for cases with $m_{1}\not=m_{2}$ and for $V_{1}(\mathbf{r})\not=V_{2}(\mathbf{r})$. At this stage it is useful to redefine the interaction coefficients as $g_{s}=\tilde{g}_{s}N_{s}$ and $g_{12}=\tilde{g}_{12}\sqrt{N_{1}N_{2}}$, obtaining TCGPEs of the form:
\begin{equation}
\label{GPE}	\left(-\frac{\hbar^{2}\nabla^{2}}{2m}+V+g_{s}|\Psi_{s}|^{2}+g_{12}\Pi^{s-3/2}|\Psi_{3-s}|^{2}-\mu_{s}\right)\Psi_{s}=0\;,
\end{equation}
with $\Pi=N_{1}/N_{2}$. 

By considering the TF limit, which neglects the kinetic energy terms ($\nabla^2\Psi_{s}=0$) when compared with the nonlinear interaction terms, we are able to write down density profiles for either component. Adopting this limit the TCGPEs [Eq.~(\ref{GPE})] become:
\begin{equation}
	\label{TF}
	\left(V+g_{s}n_{s}+g_{12}\Pi^{s-3/2}n_{3-s}-\mu_{s}\right)\Psi_{s}=0\;,
\end{equation}
where we define ${n_{s}(\mathbf{r})=|\Psi_{s}(\mathbf{r})|^{2}}$ for $s=1$ and $2$.
Then, by solving the two coupled equations (\ref{TF}) one obtains the general form of the TF density profile for each component in the region where both components coexist, i.e. $n_{s}\neq0$, for both values of $s$:
\begin{equation}
\label{nstwo}
n_{s}(\mathbf{r})=\frac{(g_{12}\Pi^{s-3/2}-g_{3-s})V(\mathbf{r})+\mu_{s}g_{3-s}-\mu_{3-s}g_{12}\Pi^{s-3/2}}{g_{1}g_{2}-g_{12}^{2}}.
\end{equation}
Note that in order to have positive-definite solutions within the TF approximation the denominator in Eq.~(\ref{nstwo}), $g_{1}g_{2}-g_{12}^{2}$, must be positive \cite{eto}. A system fulfilling this condition is commonly said to be in the miscible regime, otherwise it is in the immiscible regime. Throughout this paper, we will only consider intra- and inter-species interaction coefficients such that we are in the miscible regime. In the regions where one component is absent ($n_{s}=0$ for $s=1$ or $2$) the density profile of the other component, within the TF approximation, reads:
\begin{equation}
\label{nsone}
	n_{s}(\mathbf{r})=\frac{\mu_{s}-V(\mathbf{r})}{g_{s}}.
\end{equation}

%%%%%%%%%%%%%%%%%%%%%%%%%%%%%%%%%%%%%%%%%%%%%%%%%%%%%%%%%%%%%%%%%%%%%%%%%%%%%%%%%%%%%%%%%%%%%%%%%%%%%%%%%%%%%%%%%%%%%%%%%%
\subsection{Thomas--Fermi boundaries}
\label{sec:HP}
\begin{figure}
    \centering
    \includegraphics[width=0.5\textwidth]{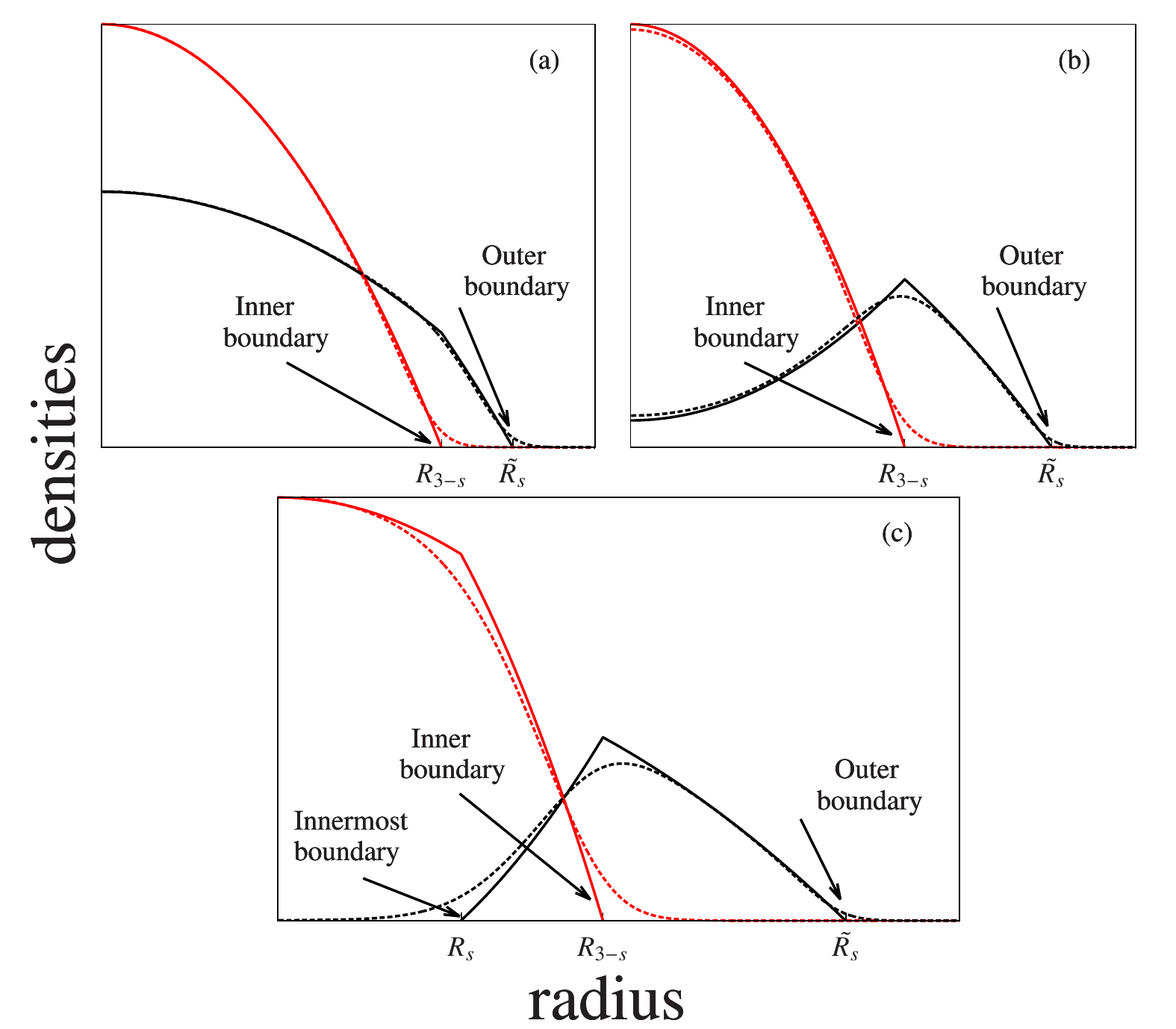}
    \caption{ Different density distributions obtained within the TF approximation (solid lines) from Eqs.~(\ref{nstwo}) and (\ref{nsone}) and by numerically integrating (dotted lines) the TCGPEs [Eq.~(\ref{GPE})] for the coexisting (a,b), and spatially separated (c) regimes. The different boundaries obtained in the TF approximation [Eqs.~(\ref{rcstworegionstwo})] are highlighted. In this plot we assume $g_{s}>g_{3-s}$ with $\Pi=1$.}
    \label{fig:fig1}
\end{figure}

In a TCBEC, the external potential and the interaction parameters determine the density distributions of the two components. In the following we consider an isotropic harmonic potential $V(\mathbf{r})=V(r)=m\omega_{r}^{2}r^{2}/2$, where $\omega_r$ is the associated angular frequency. With such a potential we can observe two different regimes: (i) the coexisting regime, where one of the components occurs only in coexistence with the other [Fig.~\ref{fig:fig1}(a,b)]; and (ii) the spatially separated regime where both components occur partly in coexistence with each other and partly in isolation [Fig.~\ref{fig:fig1}(c)]. In general, we will denote the component with largest support (meaning the component with the largest spatial extent) with the subscript $s$. Within the TF approximation, we can distinguish two cases, when $g_{3-s}> g_{12}\Pi^{s-3/2}$ (equivalently $\tilde{g}_{{3-s}}>\tilde{g}_{12}$) in which case both components have their density maxima at the center of the trap [Fig.~\ref{fig:fig1}(a)], or otherwise, when $g_{3-s}< g_{12}\Pi^{s-3/2}$ in which case component $3-s$ has its maximum of density at the centre of the trap while component $s$ has its maximum of density away from the centre [Fig.~\ref{fig:fig1}(b,c)]. One can then note that if $n_{s}(0)>0$ ($=0$) we are in the coexisting regime [Fig.~\ref{fig:fig1}(a,b)] (spatially separated regime [Fig.~\ref{fig:fig1}(c)]). The condition separating these two regimes is derived in Sec.~\ref{sec:TFG}.

The Thomas--Fermi limit, as introduced above, can provide us with a handle on the relative extent of each component. As such, we define the TF radii $R_{s}$, $R_{3-s}$ and $\tilde{R}_{s}$ of the TCBEC trapped in an external harmonic potential by imposing $n_{s}(R_{s})=0$, $n_{s}(R_{3-s})=0$ in Eqs.~(\ref{nstwo}) and $n_{s}(\tilde{R}_{s})=0$ in Eq.~(\ref{nsone}), respectively:
\begin{subequations}
	\label{rcstworegionstwo}
\begin{align}
\label{rcstworegions}
	&R_{s}^{2}=\frac{2}{m\omega_{r}^{2}}\frac{g_{12}\Pi^{s-3/2}\mu_{3-s}-g_{3-s}\mu_{s}}{g_{12}\Pi^{s-3/2}-g_{3-s}},\\
	\label{rcstworegions1}
	&R_{3-s}^{2}=\frac{2}{m\omega_{r}^{2}}\frac{g_{12}\mu_{s}-g_{s}\Pi^{s-3/2}\mu_{3-s}}{g_{12}-g_{s}\Pi^{s-3/2}},\\
	&\tilde{R}_{s}^{2}=\frac{2}{m\omega_{r}^{2}}\mu_{s}.\label{rcstworegions2}
\end{align}
\end{subequations}
In Fig.~\ref{fig:fig1} we plot the density profiles of the TCBEC within the TF approximation and indicate the three boundaries given in Eq.~(\ref{rcstworegionstwo}); the outer boundary ($\tilde{R}_{s}$) that gives the maximum extent of the BEC, and the inner ($R_{3-s}$) and innermost ($R_{s}$) boundaries delimiting the regions where the two components coexist. Note that the innermost boundary appears only when $n_s(0)=0$ (i.e.\ $R_{s}$ is only defined when we are in the spatially separated regime). Figure~\ref{fig:fig1} also shows the numerical solution of the TCGPE [Eq.~(\ref{GPE})] in all the discussed cases, demonstrating that the TF approach cannot accurately describe the density profiles close to the boundaries. In the next section we will go beyond the TF approximation, deriving universal equations governing the profiles of the densities around the TF boundaries (the low density regions), thus enabling an improvement on the approximate analytical profiles.

%%%%%%%%%%%%%%%%%%%%%%%%%%%%%%%%%%%%%%%%%%%%%%%%%%%%%%%%%%%%%%%%%%%%%%%%%%%%%%%%%%%%%%%%%%%%%%%%%%%%%%%%%%%%%%%%%%%%%%%%%%

\section{Universal equation}
\label{sec:BTF}
\subsection{Overview}
\label{sec:universaloverview}
In this section we present an analytical procedure to obtain the density profile of a TCBEC around the outer, inner, and innermost boundaries. We derive a universal equation that describes the density profile around each boundary, generalizing to two components the method developed in \cite{Stringari_1996} for the single component case. This generalization will require additional approximations for the inner and innermost boundaries, beyond those for the one component case. Note that in all the expressions shown in this section the subscript $s$ refers to the component with largest support.

%%%%%%%%%%%%%%%%%%%%%%%%%%%%%%%%%%%%%%%%%%%%%%%%%%%%%%%%%%%%%%%%%%%%%%%%%%%%%%%%%%%%%%%%%%%%%%%%%%%%%%%%%%%%%%%%%%%%%%%%%%

\subsection{Outer boundary}
\label{sec:externalboundary}
In the vicinity of the outer boundary, $\tilde{R}_{s}$ (present in both coexisting and spatially separated regimes), the TCBEC behaves effectively as if it had a single component. Therefore, we can follow the lines of \cite{Stringari_1996} and linearize the harmonic potential around $\tilde{R}_{s}$:
\begin{equation}
\label{linearization}
	V(r)\simeq V(\tilde{R}_{s})+ m\omega_{r}^{2}\tilde{R}_{s}(r-\tilde{R}_{s})+O((r-\tilde{R}_{s})^2).
\end{equation} 
We then introduce this linearization into Eq.~(\ref{GPE}) with $\Psi_{3-s}=0$, obtaining:
\begin{equation}
\label{LGPE1C}
\left[-\frac{\hbar^{2}}{2m}\frac{\partial^{2}}{\partial r^{2}}+m\omega_{r}^{2}\tilde{R}_{s}(r-\tilde{R}_{s})+{g_{s}}|\Psi_{s}|^{2}\right]\Psi_{s}=0\;,
\end{equation}
where we have used that $\mu_{s}=V(\tilde{R}_{s})$ [see Eq.~(\ref{nsone})], and we have only kept the second derivative term of the radial part of the Laplacian, i.e., in two (three) dimensions we impose 
$r^{-1}\partial\Psi/\partial r \ll \partial^{2}\Psi/\partial r^{2}$
($2r^{-1}\partial\Psi/\partial r \ll \partial^{2}\Psi/\partial r^{2}$). 
This approximation applies for values of $\tilde{R}_{s}$ much larger than the thickness of the boundary given by Eq.~(\ref{d1}), as discussed in \cite{Stringari_1996}.

 Finally, by defining the dimensionless variable
 \begin{equation}
 \label{xi1}
	\tilde{\xi_{s}}=\frac{r-\tilde{R}_{s}}{\tilde{d_{s}}},
 \end{equation} 
with
 \begin{equation}
 \label{d1}
	 \tilde{d_{s}}=\left(\frac{\hbar^{2}}{2m^{2}\omega_{r}^{2}\tilde{R}_{s}}\right)^{1/3},
 \end{equation}
and the dimensionless wave function $\phi_{s}$ through
\begin{equation}
\label{xi1psi}
	\Psi_{s}(r)=\frac{\hbar}{\tilde{d_{s}}\sqrt{2m g_{s}}}\phi_{s}(\tilde{\xi_{s}}),
\end{equation} 
we obtain the following universal equation describing the profile of the outer boundary \cite{Stringari_1996}:
\begin{equation}
\label{universal}
	\phi_{s}''-(\tilde{\xi}_{s}+\phi_{s}^{2})\phi_{s}=0.
\end{equation}

%%%%%%%%%%%%%%%%%%%%%%%%%%%%%%%%%%%%%%%%%%%%%%%%%%%%%%%%%%%%%%%%%%%%%%%%%%%%%%%%%%%%%%%%%%%%%%%%%%%%%%%%%%%%%%%%%%%%%%%%%%

 \subsection{Inner and innermost boundaries}
 \label{sec:interiorboundary}
Around the inner and innermost boundaries, $R_{s}$ and $R_{3-s}$, respectively [see Fig.~\ref{fig:fig1}(c)], both components coexist, and we must therefore  consider the full coupled Eqs.~(\ref{GPE}). Thus, in order to obtain the density profile around the $s$ component boundary (the innermost boundary) we linearize the potential around $R_{s}$:
\begin{equation}
V(r)\simeq V(R_{s})+ m\omega_{r}^{2}R_{s}(r-R_{s})+O((r-R_{s})^2).
\end{equation}
Introducing this linearization into Eqs.~(\ref{GPE}), one obtains the following two coupled equations for the $s$ and $3-s$ component, respectively:
\begin{widetext}
\begin{subequations}
\label{LCGPET}
\begin{align}
\label{LCGPE1}
&\left[-\frac{\hbar^{2}}{2m}\frac{\partial^{2}}{\partial r^{2}}+\frac{g_{12}\Pi^{s-3/2}(\mu_{3-s}-\mu_{s})}{g_{12}\Pi^{s-3/2}-g_{3-s}}+m\omega_{r}^{2}R_{s}(r-R_{s})+{g_{s}}|\Psi_{s}|^{2}+{g_{12}}\Pi^{s-3/2}|\Psi_{3-s}|^{2}\right]\Psi_{s}=0\;,\\
&\left[-\frac{\hbar^{2}}{2m}\frac{\partial^{2}}{\partial r^{2}}+\frac{g_{3-s}(\mu_{3-s}-\mu_{s})}{g_{12}\Pi^{s-3/2}-g_{3-s}}+m\omega_{r}^{2}R_{s}(r-R_{s})+{g_{3-s}}|\Psi_{3-s}|^{2}+{g_{12}}\Pi^{3/2-s}|\Psi_{s}|^{2}\right]\Psi_{3-s}=0\label{LCGPE}\;,
\end{align}
\end{subequations}
\end{widetext}
where the term $V(R_{s})$ in both equations has been rewritten using the expression (\ref{rcstworegions}). As for the outer boundary, we only keep the second derivative term of the Laplacian. The influence of the first derivative is much less than that of the second derivative in the limit of large $R_{s}$. Thus, this approximation will not be valid in the cases for which $R_{s}$ is close to the origin. Specifically, $R_s$ must be larger than the thickness of the boundary [Eq.~(\ref{d2})] \cite{Stringari_1996}, or, in other words, the relationship between the nonlinear parameters has to be such that the system is far from the crossover condition between the coexisting and spatially separated regimes, for which $R_{s}=0$ (analytical expression shown in Sec.~\ref{sec:TFG}).

In order to solve the two coupled Eqs.~(\ref{LCGPET}) for the $s$ component, we use the TF approximation for the $3-s$ component by assuming that, close to $R_{s}$, the density of component $3-s$ is large enough to ignore the kinetic energy terms, i.e., we impose $\partial^{2} \Psi_{3-s}/\partial r^{2}=0$ in Eq.~(\ref{LCGPE}). There are two limiting cases where this assumption cannot be applied: (i) when $R_{3-s}-R_{s}=O(\epsilon)$, which occurs when $g_{12}\rightarrow\sqrt{g_{1}g_{2}}$, because the TF approach is at the limit of its applicability and (ii) when $\tilde{R}_{s}-R_{3-s}=O(\epsilon)$, which occurs for $g_{12}\rightarrow0$, because there is no interaction between components and the system reduces to two noninteracting BECs.

Then, by combining the TF form of Eq.~(\ref{LCGPE}) with Eq.~(\ref{LCGPE1}), one obtains:
 \begin{align}
 \label{GPETF}
\left[-\frac{\hbar^{2}}{2m}\frac{\partial^{2}}{\partial r^{2}}\right.&+m\omega_{r}^{2}R_{s}\left(1-\frac{g_{12}\Pi^{s-3/2}}{g_{3-s}}\right)\left(r-R_{s}\right)\nonumber\\
+&\left.\left(g_{s}-\frac{g_{12}^{2}}{g_{3-s}}\right)|\Psi_{s}|^{2}\right]\Psi_{s}=0\;.
 \end{align}
By following a similar procedure one obtains the equivalent equation for the $3-s$ component around $R_{3-s}$ at which $n_{3-s}(R_{3-s})=0$. The resulting equation has the same form as Eq.~(\ref{GPETF}) but with $3-s$ and $s$ exchanged. Thus, in order to solve the innermost and inner boundaries we define the dimensionless variable:
 \begin{equation}
 \label{xi2}
	 \xi_{s}=\pm\frac{r-R_{s}}{d_{s}},
 \end{equation} 
with $d_{s}$ given by
 \begin{equation}
 \label{d2}
	 d_{s}=\left[\frac{\hbar^{2}}{\pm 2m^{2}\omega_{r}^{2}R_{s}\left(1-g_{12}\Pi^{s-3/2}/g_{3-s}\right)}\right]^{1/3},
 \end{equation}
and the dimensionless wave function $\phi_s$ defined through
\begin{equation}
\label{xi2psi}
	\Psi_{s}(r)=\frac{\hbar}{d_{s}\sqrt{2m \left(g_{s}-g_{12}^{2}/g_{3-s}\right)}}\phi_s(\xi_{s}).
\end{equation} 
In Eqs.~(\ref{xi2}) and (\ref{d2}) the $+$ ($-$) sign applies for the inner (innermost) boundary, and for the inner boundary $s$ has to be interchanged by $3-s$ and $3-s$ by $s$.
 
Finally, one obtains the same universal equation derived in Sec. \ref{sec:externalboundary} [Eq.~(\ref{universal})] for both boundaries:
\begin{equation}
\label{universal1}
	\phi_s''-(\xi_s+\phi_s^{2})\phi_s=0,
\end{equation}
where the $s$ ($3-s$) applies for the innermost (inner) boundary.

Summarizing, in this section we have obtained the universal Eqs.~(\ref{universal}) and (\ref{universal1}) that describe the density profiles at the boundaries of a TCBEC trapped in a harmonic potential. Note that we obtain the same universal equation for the three boundaries. However, they require different transformations, Eqs.~(\ref{xi1}) and (\ref{xi1psi}) for the outer boundary and Eqs.~(\ref{xi2}) and (\ref{xi2psi}) for the inner and innermost boundaries, in order to retrieve the actual wave function at each boundary. 
%%%%%%%%%%%%%%%%%%%%%%%%%%%%%%%%%%%%%%%%%%%%%%%%%%%%%%%%%%%%%%%%%%%%%%%%%%%%%%%%%%%%%%%%%%%%%%%%%%%%%%%%%%%%%%%%%%%%%%%%%%

\subsection{Solving the universal equation}\label{sec:universal}
One can recognize Eqs.~(\ref{universal}) and (\ref{universal1}) in the literature \cite{HM_1980,Segur_1981} as being a Painleve type-II equation, which for positive defined solutions with no divergences or sinusoidal behaviors has a Hastings--McLeod (HM) solution\footnotemark[1] \cite{HM_1980} with the following asymptotics:
\footnotetext[1]{Note that the Painleve type-II equation has a factor two in front of the nonlinear term that makes the prefactor in front of the Hastings--McLeod solution equal to 1 instead of $\sqrt{2}$.}
\begin{equation}
	\label{HM}
 \phi_{\textrm{HM}}(\xi) \sim
  \begin{cases}
   \sqrt{2}\text{Ai}(\xi) & \text{for } \xi \rightarrow +\infty, \\
   \sqrt{-\xi}       & \text{for } \xi \rightarrow -\infty,
  \end{cases}	
\end{equation} 
where $\text{Ai}(\xi)$ is the Airy function.

  \begin{figure}
    \centering
    \includegraphics[width=0.5\textwidth]{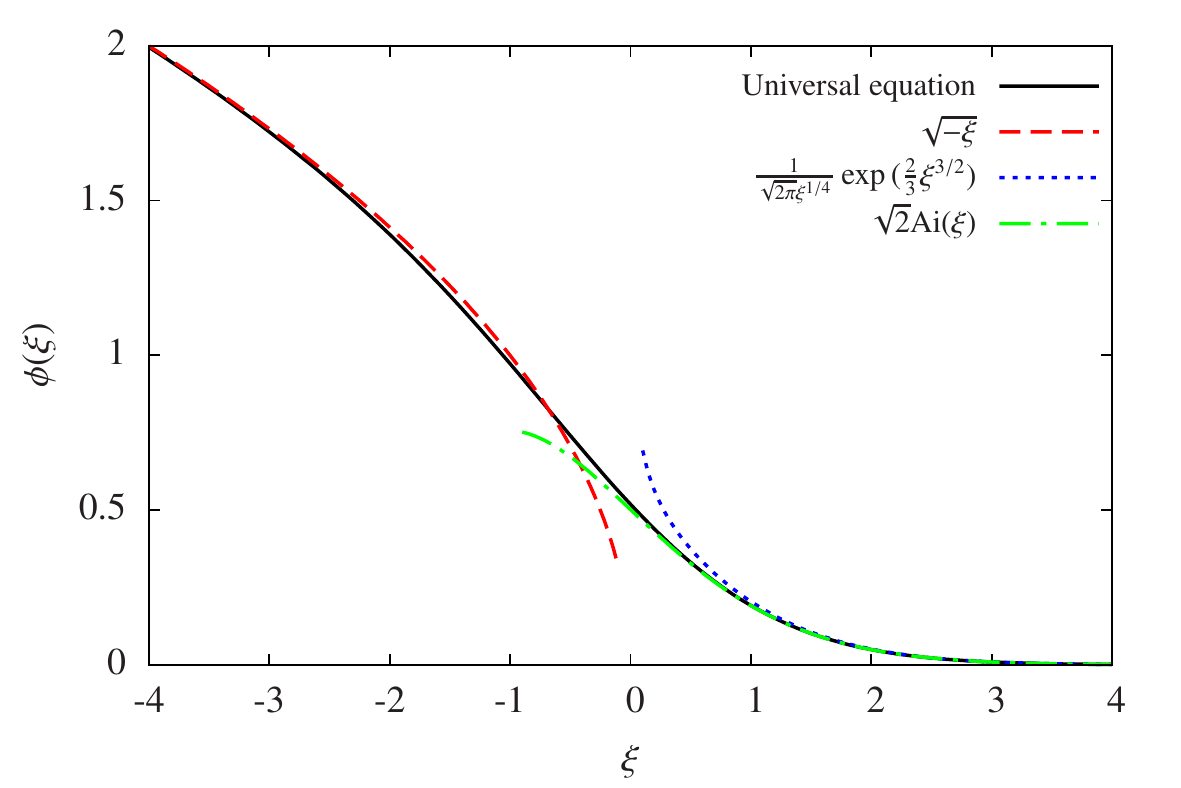}
    \caption{Numerical solution of the universal equation [Eq.~(\ref{universal}) and (\ref{universal1})] (black solid line), and the two asymptotic behaviors given by the Hastings--McLeod solution [Eq.~(\ref{HM})]: the Airy function (green dotted dashed line), and $\sqrt{-\xi}$ (red dashed line). We also plot the asymptotic behavior of the Airy function [Eq.~(\ref{airy1})] for $\xi\rightarrow+\infty$ (blue dotted line).}
    \label{fig:universal}
\end{figure}

In this paper we are interested in the $\xi\rightarrow\infty$ limit, where the density is small and the TF approach is not applicable. In this limit, the asymptotic formula of the Airy function gives an intuitive form of the behavior of the wave function at large $\xi$:
\begin{equation}
\label{airy1}
	\phi(\xi\rightarrow+\infty)\simeq\sqrt{2}\text{Ai}(\xi)\simeq \frac{1}{\sqrt{2\pi}\xi^{1/4}}e^{-2\xi^{3/2}/3}.
\end{equation}
Figure~\ref{fig:universal} shows the asymptotic forms of the HM solution [Eq.~(\ref{HM})] and the numerical solution of Eq.~(\ref{universal}) or (\ref{universal1}) for comparison. The asymptotic behavior of the Airy function [Eq.~(\ref{airy1})] is also plotted. Even though the approximation given by Eq.~(\ref{airy1}) is obtained at $\xi\rightarrow\infty$, the two functions coincide even for low values of $\xi$, which makes this asymptotic approximation very useful as an analytical expression to describe the universal equation for $\xi>0$. It remains for us to find appropriate expressions for the TF boundaries in order to complete the transformations to retrieve the actual wave functions near each boundary.
%%%%%%%%%%%%%%%%%%%%%%%%%%%%%%%%%%%%%%%%%%%%%%%%%%%%%%%%%%%%%%%%%%%%%%%%%%%%%%%%%%%%%%%%%%%%%%%%%%%%%%%%%%%%%%%%%%%%%%%%%%

%%%%%%%%%%%%%%%%%%%%%%%%%%%%%%%%%%%%%%%%%%%%%%%%%%%%%%%%%%%%%%%%%%%%%%%%%%%%%%%%%%%%%%%%%%%%%%%%%%%%%%%%%%%%%%%%%%%%%%%%%%
\section{Thomas--Fermi solutions}
\label{sec:TFG}
\subsection{Overview}
\label{sec:thomasfermioverview}
In this section we present a general procedure to obtain the radii and chemical potentials within the TF approximation of a TCBEC trapped in an isotropic harmonic potential in 1D, 2D and 3D. We study both the coexisting and the spatially separated regimes, and we determine the frontier between them for each dimensionality.

The TF solution, either in the coexisting or in the spatially separated regime, can be found by following three general steps (a similar procedure was presented in \cite{Shenoy_1996}): (i) normalization of the density using the proper limits of integration, (ii) isolation of the chemical potential of each of the components, $\mu_{s}$ and $\mu_{3-s}$, as a function of the parameters of the system from the normalization integrals and (iii) insertion of $\mu_{s}$ and $\mu_{3-s}$ into Eq.~(\ref{nstwo}) [Eq.~(\ref{nsone})] for the coexisting (non-coexisting) region, to obtain the density profile.

In some cases, these steps can be laborious, and in particular step (iii) may not be available analytically. Specifically, we have found that in the 1D and 3D cases the chemical potentials cannot be inverted analytically within the spatially separated regime. Thus, here we show how to reduce the two coupled algebraic equations to a single equation, reducing the complexity of the numerical problem \cite{Tsubota_2001,Modugno_2002}. In all other cases, fully analytical expressions can be found.
%%%%%%%%%%%%%%%%%%%%%%%%%%%%%%%%%%%%%%%%%%%%%%%%%%%%%%%%%%%%%%%%%%%%%%%%%%%%%%%%%%%%%%%%%%%%%%%%%%%%%%%%%%%%%%%%%%%%%%%%%%
\comment{\subsection{General definitions}
We also define the parameters that appear in the normalization integrals: $\alpha_{\text{1}},\;\beta_{\text{1}},\;\gamma_{\text{1}},\;\Omega_{\text{1}}\text{ and }\epsilon_{\text{1}}$:
\begin{equation}
\alpha_{\text{1}}=-\frac{2m\omega_{r}^{2}}{3}\frac{g_{12}}{g_{3-s}}\left(\frac{g_{12}-g_{3-s}\Pi^{3/2-s}}{g_{1}g_{2}-g_{12}^{2}}\right)\nonumber
\end{equation}\vspace{-0.5cm}
\begin{equation}
\label{alpha_beta_gamma}
\begin{aligned}[c]
\beta_{\text{1}}&=-\frac{2m\omega_{r}^{2}}{3}\frac{g_{12}\Pi^{3/2-s}-g_{s}}{g_{1}g_{2}-g_{12}^{2}}\\
\Omega_{\text{1}}&=\frac{2m\omega_{r}^{2}}{3}\frac{g_{12}}{g_{s}}\left(\frac{g_{12}-g_{s}\Pi^{s-3/2}}{g_{1}g_{2}-g_{12}^{2}}\right)
\end{aligned}\;
\begin{aligned}[c]
\epsilon_{\text{1}}&=\frac{2m\omega_{r}^{2}}{3}\frac{1}{g_{s}}\\
\gamma_{\text{1}}&=\frac{2m\omega_{r}^{2}}{3}\frac{g_{12}\Pi^{s-3/2}-g_{3-s}}{g_{1}g_{2}-g_{12}^{2}}
\end{aligned}
\end{equation}
Note that $\alpha_{\text{1}}$ and $\beta_{\text{1}}$ are related with $\gamma_{\text{1}}$ and $\Omega_{\text{1}}$ by ${\alpha_{\text{1}}=-\frac{g_{12}\Pi^{3/2-s}}{g_{3-s}}\gamma_{\text{1}}}$ and $\beta_{\text{1}}=-\frac{g_{s}}{g_{12}\Pi^{s-3/2}}\Omega_{\text{1}}$.}\\

%%%%%%%%%%%%%%%%%%%%%%%%%%%%%%%%%%%%%%%%%%%%%%%%%%%%%%%%%%%%%%%%%%%%%%%%%%%%%%%%%%%%%%%%%%%%%%%%%%%%%%%%%%%%%%%%%%%%%%%%%%
\subsection{Coexisting regime}
The coexisting regime can be solved in a fully analytical fashion for all three dimensionalities. In this case, the normalization conditions for the $s$ and $3-s$ components read:
\begin{subequations}
\begin{align}
\label{tffirst}
&\int\limits^{R_{3-s}}_{0}{(\zeta_{s}+\eta_{s}r^{2})\,d^{D}r}+\!\int\limits^{\tilde{R}_{s}}_{R_{3-s}}{(\lambda_{s}+\kappa_{s}r^{2})\,d^{D}r}=1,\\
&\int\limits^{R_{3-s}}_{0}{(\zeta_{3-s}+\eta_{3-s}r^{2})\,d^{D}r}=1,\label{tffirst1}
\end{align}
\end{subequations}
where $D=1$, 2 or 3 depending on the dimensionality. The $d^{D}r$ differential represents the volume element for each case: ${d^{1}r=2dr,\;d^{2}r=2\pi rdr,\;d^{3}r=4\pi r^{2}dr}$, (note that in the 1D case we add a factor 2 due to the fact that $r$ should go from $-\infty$ to $\infty$). We assume cylindrically and spherically isotropic configurations in the 2D and 3D harmonic potentials, respectively, and $\lambda_{s}$, $\kappa_{s}$, $\zeta_{s}$ and $\eta_{s}$ are given by:
\begin{equation}
\label{lambda_zeta_eta}
\begin{aligned}[c]
\lambda_{s}&=\frac{\mu_{s}}{g_{s}},\\
\kappa_{s}&=-\frac{m\omega_{r}^{2}}{2g_{s}},
\end{aligned}
\qquad\qquad
\begin{aligned}[c]
\zeta_{s}&=\frac{\mu_{s}g_{3-s}-\mu_{3-s}g_{12}\Pi^{s-3/2}}{g_{1}g_{2}-g_{12}^{2}},\\
\eta_{s}&=\frac{m\omega_{r}^{2}}{2}\frac{g_{12}\Pi^{s-3/2}-g_{3-s}}{g_{1}g_{2}-g_{12}^{2}}.
\end{aligned}
\end{equation}
By carrying out the integrations in Eq.~(\ref{tffirst}) and (\ref{tffirst1}) we reach:
\begin{subequations}
	\label{TFG0}
\begin{align}
\label{TFG1}
	\Omega_{D} R_{3-s}^{D+2}+\epsilon_{D} &\tilde{R}_{s}^{D+2}=1,\\
	\beta_{D} R_{3-s}^{D+2}&=1,\label{TFG2}
	\end{align}
\end{subequations}
where $\Omega_{D}$, $\epsilon_{D}$ and $\beta_{D}$ read:
\begin{equation}
\label{alpha_beta_gamma1}
\begin{aligned}
\beta_{D}&=-\Delta_{D}\frac{2m\omega_{r}^{2}}{3}\frac{g_{12}\Pi^{3/2-s}-g_{s}}{g_{1}g_{2}-g_{12}^{2}},\\
\Omega_{D}&=\Delta_{D}\frac{2m\omega_{r}^{2}}{3}\frac{g_{12}}{g_{s}}\left(\frac{g_{12}-g_{s}\Pi^{s-3/2}}{g_{1}g_{2}-g_{12}^{2}}\right),\\
\epsilon_{D}&=\Delta_{D}\frac{2m\omega_{r}^{2}}{3}\frac{1}{g_{s}},
\end{aligned}
\end{equation}
and where the scaling factors
\begin{equation}
\label{eq:Delta}
\Delta_{1} = 1,
\qquad
\Delta_{2} = \frac{3\pi}{8},
\qquad
\Delta_{3} =  \frac{6\pi}{15},
\end{equation}
account for the different dimensionalities. Note that $\beta_{D}$ is related to $\Omega_{D}$ through $\beta_{D}=-(g_{s}/g_{12}\Pi^{s-3/2})\Omega_{D}$.

Then, using Eqs.~(\ref{TFG0}) and the definitions of Eqs.~(\ref{rcstworegionstwo}) we obtain:
\begin{subequations}
\begin{align}
 \label{muscr}
\mu_{s}&=\frac{m\omega_{r}^{2}}{2\epsilon_{D}^{2/(D+2)}}\left(1-\Omega_{D}/\beta_{D}\right)^{2/(D+2)},\\
 \label{mu3scr}
	\mu_{3-s}&=\frac{g_{12}\mu_{s}}{\Pi^{s-3/2}  g_{s}}+\frac{m \omega_{r}^{2}}{2}\left(\frac{g_{s}-\Pi^{3/2-s}g_{12}}{g_{s}}\right)\beta_{D}^{-2/(D+2)}.
\end{align}
\end{subequations}
Finally, by introducing these two chemical potentials into Eq.~(\ref{nstwo}) and Eq.~(\ref{nsone}) one finds the solution of the TF density profile of a TCBEC in the coexisting regime.

The analytical expression within the TF approximation of the frontier between the coexisting and spatially separated regimes can be found by using the fully analytical expression of the density profile obtained by inserting Eq.~(\ref{muscr}) and Eq.~(\ref{mu3scr}) into Eq.~(\ref{nstwo}) and setting $n_{s}(0)=0$:
\begin{align}
\label{eq:condition}
g_{3-s}=\frac{g_{12}^2}{g_{s}} -g_{12}(g_{12}-\Pi^{s-3/2}  g_{s})\left(\frac{g_{s}+\Pi^{s-3/2}  g_{12}}{g_{s}^{-D/2} \Pi^{s-3/2} g_{12}}\right)^{-2/D}.
\end{align}
Note that Eq.~(\ref{eq:condition}) gives the condition that separates coexisting [Fig.~\ref{fig:fig1}(a,b)] and spatially separated [Fig.~\ref{fig:fig1}(c)] regimes. All the calculations shown in this subsection are valid for $D=1$, 2 and 3.
%%%%%%%%%%%%%%%%%%%%%%%%%%%%%%%%%%%%%%%%%%%%%%%%%%%%%%%%%%%%%
%%%%%%%%%%%%%%%%%%%%%%%%%%%%%%%%%%%%%%%%%%%%%%%%%%%%%%%%%%%%%%%%%%%%%%%%%%%%%%%%%%%%%%%%%%%%%%%%%%%%%%%%%%%%%%%%%%%%%%%%%%
\subsection{Spatially separated regime}
As mentioned previously, the full solution of the two chemical potentials in the spatially separated regime cannot be found analytically in the one and three dimensional cases. Here we present a procedure to reduce the complexity of this numerical problem. We start by using the normalization conditions for the $s$ and $3-s$ components, respectively:
\begin{subequations}
\label{spsrint}
\begin{align}
&\int\limits^{R_{3-s}}_{R_{s}}{(\zeta_{s}+\eta_{s}r^{2})\,d^{D}r}+\int\limits^{\tilde{R}_{s}}_{R_{3-s}}{(\lambda_{s}+\kappa_{s}r^{2})\,d^{D}r}=1,\\
&\int\limits^{R_{s}}_{0}{(\lambda_{3-s}+\kappa_{3-s}r^{2})\,d^{D}r}+\int\limits^{R_{3-s}}_{R_{s}}{(\zeta_{3-s}+\eta_{3-s}r^{2})\,d^{D}r}=1,
\end{align}
\end{subequations}
where $\lambda_{s}$, $\kappa_{s}$, $\zeta_{s}$ and $\eta_{s}$ are defined in Eq.~(\ref{lambda_zeta_eta}).
After integrating Eqs.~(\ref{spsrint}) and rearranging the terms we obtain the two coupled equations:
\begin{subequations}
\label{TFG}
\begin{align}
		\gamma_{D} &R_{s}^{D+2}+\Omega_{D} R_{3-s}^{D+2}+\epsilon_{D} \tilde{R}_{s}^{D+2}=1,\\
		&\alpha_{D} R_{s}^{D+2}+\beta_{D} R_{3-s}^{D+2}=1,
\end{align}
\end{subequations}
where $\alpha_{D}$ and $\gamma_{D}$ read:
\begin{align}
\label{alpha_beta_gamma2}
\alpha_{D}&=-\Delta_{D}\frac{2m\omega_{r}^{2}}{3}\frac{g_{12}}{g_{3-s}}\left(\frac{g_{12}-g_{3-s}\Pi^{3/2-s}}{g_{1}g_{2}-g_{12}^{2}}\right),\nonumber\\
\gamma_{D}&=\Delta_{D}\frac{2m\omega_{r}^{2}}{3}\frac{g_{12}\Pi^{s-3/2}-g_{3-s}}{g_{1}g_{2}-g_{12}^{2}},
\end{align}
with the scaling factors accounting for the different dimensionalities given in Eq.~(\ref{eq:Delta}), and where 
$\Omega_{D}$, $\epsilon_{D}$ and $\beta_{D}$ are defined as in Eq.~(\ref{alpha_beta_gamma1}).

In the 1D and 3D cases the chemical potentials of both components cannot be obtained analytically from Eqs.~(\ref{TFG}). Therefore in order to reduce the two coupled equations into a single equation we rewrite $R_{s}^{D+2}$ and $R_{3-s}^{D+2}$ from Eq.~(\ref{TFG}) as:
\begin{subequations}
\label{rcsolutionsg}
\begin{align}
\label{rcsolutionsg1}
R_{s}^{D+2}&=\frac{g_{3-s}}{(g_{1}g_{2}-g_{12}^{2})\gamma_{D}}\left(g_{s}N'+g_{12}\Pi^{s-3/2}\right),\\
R_{3-s}^{D+2}&=\frac{g_{12}}{(g_{12}^{2}-g_{1}g_{2})\Omega_{D}}\left(g_{12}N'+g_{3-s}\Pi^{s-3/2}\right),\label{rcsolutionsg2}
\end{align}
\end{subequations}
where $N'=1-\epsilon_{D} \tilde{R}_{s}^{D+2}$. Introducing Eqs.~(\ref{rcsolutionsg}) into the relation between the TF radii obtained from Eqs.~(\ref{rcstworegionstwo}) yields
\begin{equation}
\label{rcrelationg}	\tilde{R}_{s}^{2}=\frac{g_{s}(g_{3-s}-g_{12}\Pi^{s-3/2})}{g_{1}g_{2}-g_{12}^{2}}R_{s}^{2}+\frac{g_{12}(g_{s}\Pi^{s-3/2}-g_{12})}{g_{1}g_{2}-g_{12}^{2}}R_{3-s}^{2},
\end{equation}
and one obtains a single equation that only depends on $\mu_{s}$ and on the parameters of the system. This equation has two roots that need to be inverted numerically in order to find $\mu_{s}$ (the root power depends on the dimensionality). By using the expression for $R_{3-s}$ from Eqs.~(\ref{rcsolutionsg1}) and (\ref{rcstworegions}) we find an analytical formulation for $\mu_{3-s}$ as a function of $\mu_{s}$.
Once we have the two chemical potentials, we introduce them into the densities [Eqs.~(\ref{nstwo}) and (\ref{nsone})], obtaining the TF solution of a TCBEC in $D$ dimensions in the spatially separated regime.

In the 2D case, however, Eqs.~(\ref{TFG}) can be solved analytically using the definitions of $\alpha,\;\beta,\;\gamma,\;\Omega$ and $\epsilon$ from Eq.~(\ref{alpha_beta_gamma1}), Eq.~(\ref{eq:Delta}) and Eq.~(\ref{alpha_beta_gamma2}) for $D=2$. In this case, using the expressions of the TF radii from Eq.~(\ref{rcstworegionstwo}) we obtain:
\begin{subequations}\label{TFmu3s}
\begin{align}
\mu_{3-s}^{\text{(2D)}}&=\sqrt{\frac{m\omega_{r}^{2}g_{3-s}(\Pi^{3-2s}+1)}{\pi}},\\
	\mu^{\text{(2D)}}_{s}&=\mu_{3-s}+\sqrt{\frac{ m\omega_{r}^{2}  (\Pi^{s-3/2}  g_{s}-g_{12}) (\Pi^{s-3/2}  g_{12}-g_{3-s})}{\pi  \Pi^{2s-3}  g_{12}}},
\end{align}
\end{subequations}
and by substituting into Eq.~(\ref{nstwo}) and Eq.~(\ref{nsone}) we get the density profiles of the TCBEC in 2D in the spatially separated regime.
%%%%%%%%%%%%%%%%%%%%%%%%%%%%%%%%%%%%%%%%%%%%%%%%%%%%%%%%%%%%%%%%%%%%%%%%%%%%%%%%%%%%%%%%%%%%%%%%%%%%%%%%%%%%%%%%%%%%%%%%%%

\begin{figure*}
\centering
    \includegraphics[width=1\textwidth]{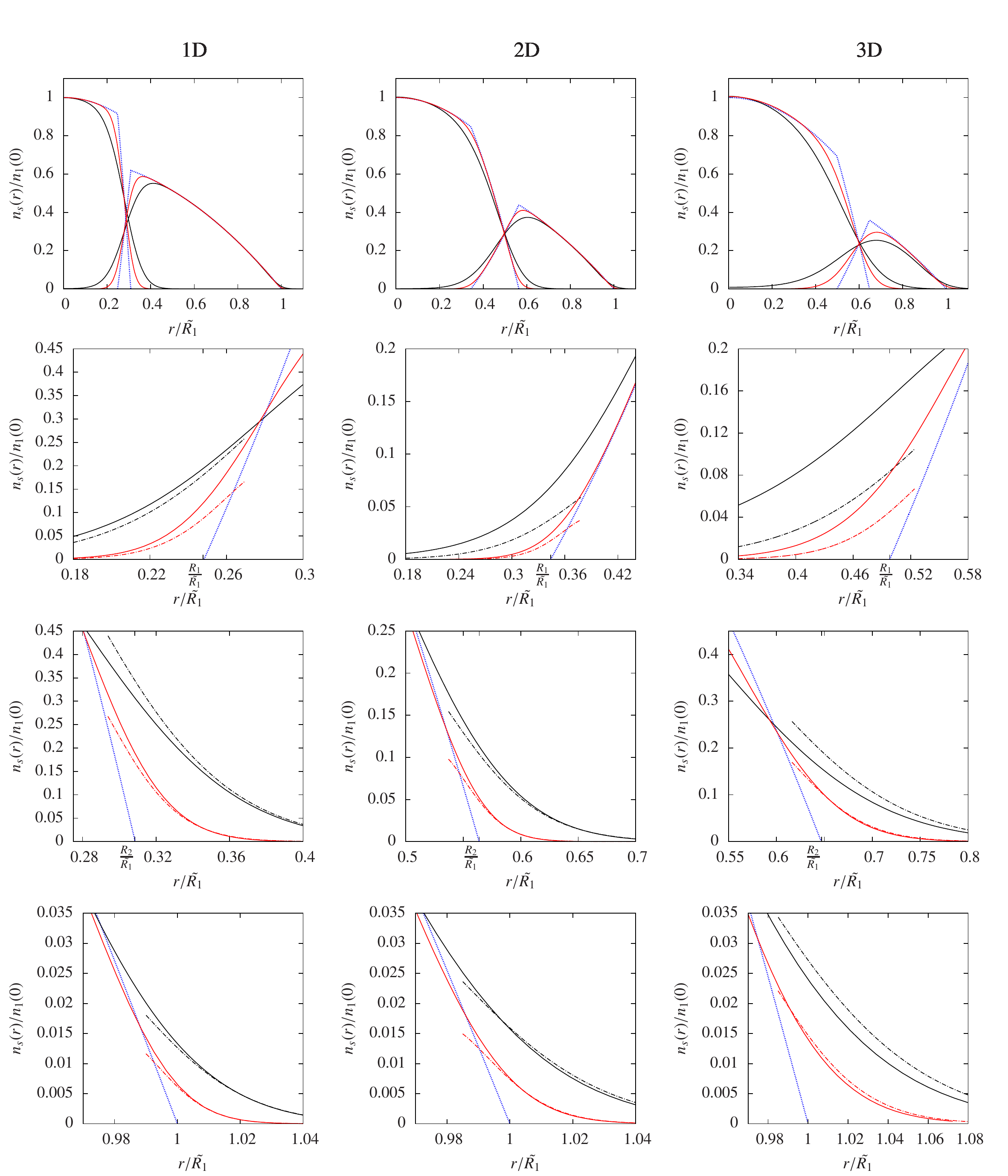}
    \caption{Comparison between the results of the density profiles obtained through the universal equation [Eqs.~(\ref{universal}) and (\ref{universal1})], the TF approximation (Sec.~\ref{sec:TFG}) and the numerical solution of the TCGPEs [Eq.~(\ref{GPE})]. The first row shows the TF density profile (blue dotted line) and the numerical solution of the TCGPEs (solid lines) of a TCBEC in the spatially separated regime for the 1D (1st column), 2D (2nd column) and 3D (3rd column) cases for two different values of $g_{2}$ in each plot. We also plot a magnification around $R_{1}$ (2nd row), $R_{2}$ (3rd row) and $\tilde{R}_{1}$ (4th row) where we include the asymptotic approximation of the universal equation $\sqrt{2}\text{Ai}(r)$ (dotted-dashed line) derived in Sec.~\ref{sec:BTF}. The parameter values used are: (i) for the 1D case $\chi=0.985$, $g_{2}=200$ (black lines) and $g_{2}=1000$ (red/gray lines), (ii) for the 2D case $\chi=0.9$, $g_{2}=1000$ (black lines) and $g_{2}=10000$ (red/gray lines) and (iii) for the 3D case $\chi=0.9$, $g_{2}=1000$ (black lines) and $g_{2}=10000$ (red/gray lines). Note that the magnifications only include the component under study. For a complete description of the dimensionless parameters and scalings see the text.}
    \label{fig:big}
\end{figure*}

%%%%%%%%%%%%%%%%%%%%%%%%%%%%%%%%%%%%%%%%%%%%%%%%%%%%%%%%%%%%%%%%%%%%%%%%%%%%%%%%%%%%%%%%%%%%%%%%%%%%%%%%%%%%%%%%%%%%%%%%%%
\section{Density profiles around the boundaries}
 \label{sec:results}
In this section, we present the comparison between the density profiles in 1D, 2D and 3D of a TCBEC around the boundaries defined in Eq.~(\ref{rcstworegionstwo}) obtained: (i) within the TF approximation (Sec.~\ref{sec:TFG}), (ii) using the universal equation derived in Sec.~\ref{sec:BTF} and (iii) by numerically integrating the TCGPEs [Eq.~(\ref{GPE})] for the spatially separated regime. To reduce the parameter phase space we consider the same number of atoms for both components, i.e., $\Pi=1$, $g_{12}=\sqrt{\chi g_{1}g_{2}}$ and $g_{s}=2g_{3-s}$, in such a way that the parameter $\chi$ determines the ratio between the different TF radii [Eq.~(\ref{rcstworegionstwo})], and $g_{3-s}$ determines the strength of the nonlinear interactions. In addition, we rescale the densities of both components to the maximum value of the density of the $3-s$ component at the origin (considering the $s$ component to be that with largest support) and the $r$ coordinate to the maximum extension of the TCBEC ($\tilde{R}_{s}$). We also use harmonic oscillator units, which is equivalent to setting $\hbar=m=\omega_{r}=1$. These settings allow us to compare the behavior, graphically, of the density profiles close to the boundaries for different values of the nonlinearity on the same axis scale.

In order to have some reference values of the strength of the considered nonlinear interactions ($g_{3-s}$) we compare our rescaled nonlinear parameters with typical experimental values. We consider a TCBEC of $^{87}\text{Rb}$ trapped in a harmonic potential with a radial trapping frequency ${\omega_{r}=2\pi\times 20\;\text{Hz}}$, transverse trapping frequency ${\omega_{\bot}=2\pi\times150\;\text{Hz}}$ (for the one and two dimensional cases) and ${a_{s}=100a_{0}}$, with $a_{0}$ and $a_{s}$ being the Bohr radius and the s-wave scattering length, respectively. In the 1D case, $g_{3-s}\simeq200$ corresponds to a BEC with ${N_{s}=N_{3-s}\simeq1\times10^{4}\;\text{particles}}$, while in the 2D and 3D cases, $g_{3-s}=1000$ corresponds to an approximate value of ${N_{s}=N_{3-s}\simeq5\times10^{4}\;\text{particles}}$.
%%%%%%%%%%%%%%%%%%%%%%%%%%%%%%%%%%%%%%%%%%%%%%%%%%%%%%%%%%%%%%%%%%%%%%%%%%%%%%%%%%%%%%%%%%%%%%%%%%%%%%%%%%%%%%%%%%%%%%%%%%

Figure~\ref{fig:big} shows the density profile of a 1D, 2D and 3D TCBEC trapped in an isotropic harmonic potential in the spatially separated regime using the TF approximation [Eqs.~(\ref{TFG})--(\ref{rcrelationg})], the universal equation [Eqs.~(\ref{universal}) and (\ref{universal1})] and the results of the TCGPE [Eq.~(\ref{GPE})] for different nonlinearities. Note that we have fixed $s=1$ for the component with largest support. We observe that in 1D (1st column of Fig.~\ref{fig:big}) the asymptotic behavior of the universal equation at the different boundaries is in excellent agreement with the numerical solution of the TCGPE, even for values of $g_{2}$ corresponding to a relatively small number of particles in a typical experimental TCBEC.

In 2D (2nd column of Fig.~\ref{fig:big}) we can see that the universal equation around the boundaries (Sec.~\ref{sec:BTF}) gives a good insight of the numerical solution of the TCGPE in a fully analytical way for $g_{2}=1\times10^{4}$. However, for relatively small nonlinearities ($g_{2}\sim1000$), the universal equation close to the innermost boundary cannot describe the density of the TCBEC, as discussed in Sec.~\ref{sec:BTF}.

In 3D (3rd column of Fig.~\ref{fig:big}) we see that the outer and inner boundaries are in very good agreement for both nonlinearities, showing that this approximation can be used to describe the density of a TCBEC around the boundaries provided that the conditions mentioned in Sec.~\ref{sec:BTF} are fulfilled. The universal equation close to the innermost boundary, on the other hand, hardly reproduces the density of the TCBEC for low values of $g_{2}$, however, the approximation appears to have broad validity for values of $g_{2}$ above $10000$.

%%%%%%%%%%%%%%%%%%%%%%%%%%%%%%%%%%%%%%%%%%%%%%%%%%%%%%%%%%%%%%%%%%%%%%%%%%%%%%%%%%%%%%%%%%%%%%%%%%%%%%%%%%%%%%%%%%%%%%%%%%
\section{Conclusions and remarks}
\label{sec:conclusions}
In this paper we have presented an analytical approximation to the ground state density profiles of a TCBEC trapped in an isotropic harmonic potential in the mean field approximation around the boundaries of each component, where the TF approximation is no longer valid. We have derived universal equations that give a very good estimation of the behavior of the density profile at the boundaries of each species, softening the sharp edges produced by the TF approximation. We have compared our analytical results with the numerically integrated TCGPE, obtaining an excellent agreement between them.
The method proposed in this paper also offers the possibility to calculate analytically, as proposed in \cite{Stringari_1996,Feder_1998}, the kinetic energy of the system, tunneling between double well potentials and other possibilities such as calculating an equivalent \textit{healing length\/} in a TCBEC in the miscible regime, similarly to the penetration depth defined in the immiscible regime by \cite{Chui_1998}. Moreover, the approach presented in this work can be easily extended to different species (i.e. different masses), and also, due to its generality, the procedure may be extended to other trapping potentials.

We have also studied the TF approximation for 1D, 2D and 3D. We have shown that the coexisting regime can be treated analytically in all three cases. However, the spatially separated regime only has analytical solution in 2D. In 1D and 3D we can decrease the complexity of the numerical inversion required by reducing the resulting system of two coupled equations to a single one. Finally, within the TF approximation, we have determined, analytically, the frontier between the coexisting and spatially separated regimes.

\begin{acknowledgments}
J. Polo and V. Ahufinger would like to thank Jordi Mompart for useful discussions and feedback. We acknowledge support from the Spanish Ministry of Economy and Competitiveness under contract FIS2011-23719 and from the Catalan Government under contract SGR2014-1639. J. Polo also acknowledges financial support from the FPI grant with reference BES-2012-053447 and from the mobility grant EEBB-I-14-08515. P. Mason and S. A. Gardiner thank the UK  EPSRC (Grant No.\ EP/K030558/1) and the European Commission for support through the Marie Curie Fellowship NUM2BEC (Grant No.\ 300285). T. P. Billam acknowledges support from the John Templeton Foundation via the Durham Emergence Project (http://www.dur.ac.uk/emergence).
\end{acknowledgments}

\bibliographystyle{apsrev4-1}

\bibliography{shorttitles,two_component}

\end{document}